# Measuring baryon acoustic oscillations with future SKA surveys

**Philip Bull**[*,1]**, Stefano Camera**[2,3]**, Alvise Raccanelli**[4,5,6]**, Chris Blake**[7]**, Pedro G. Ferreira**[8]**, Mario G. Santos**[9,10]**, Dominik J. Schwarz**[11]

[1]*Institute of Theoretical Astrophysics, University of Oslo, PO Boks 1029 Blindern, 0315 Oslo, Norway;* [2]*Jodrell Bank Centre for Astrophysics, University of Manchester, Manchester M13 9PL, UK;* [3]*CENTRA, Instituto Superior Tecnico, Universidade de Lisboa, Av. Rovisco Pais 1, 1049-001, Lisboa, Portugal;* [4]*Department of Physics & Astronomy, Johns Hopkins University, 3400 N. Charles St., Baltimore, MD 21218, USA;* [5]*Jet Propulsion Laboratory, California Institute of Technology, Pasadena CA 91109, USA;* [6]*California Institute of Technology, Pasadena CA 91125, USA;* [7]*Centre for Astrophysics & Supercomputing, Swinburne University of Technology, PO Box 218, Hawthorn, VIC 3122, Australia;* [8]*Astrophysics, University of Oxford, UK;* [9]*Physics Department, University of the Western Cape, Cape Town 7535, South Africa;* [10]*SKA SA, 3rd Floor, The Park, Park Road, Pinelands, 7405, South Africa;* [11]*Fakultät für Physik, Universität Bielefeld, Postfach 100131, D-33501 Bielefeld, Germany*
*E-mail:* p.j.bull@astro.uio.no

The imprint of baryon acoustic oscillations (BAO) in large-scale structure can be used as a standard ruler for mapping out the cosmic expansion history, and hence for testing cosmological models. In this chapter we briefly describe the scientific background to the BAO technique, and forecast the potential of the Phase 1 and 2 SKA telescopes to perform BAO surveys using both galaxy catalogues and intensity mapping, assessing their competitiveness with current and future optical galaxy surveys. We find that a 25,000 deg$^2$ intensity mapping survey on a Phase 1 array will preferentially constrain the radial BAO, providing a highly competitive 2% constraint on the expansion rate at $z \simeq 2$. A 30,000 deg$^2$ galaxy redshift survey on SKA2 will outperform all other planned experiments for $z \lesssim 1.4$.

[*]Speaker.

## 1. Introduction

The Baryon Acoustic Oscillations (BAO) are a relic from the time when photons and baryons were coupled in the early Universe, and constitute a preferred clustering scale in the distribution of matter on cosmological scales. Since the physical scale of the oscillations can be inferred from observations of the cosmic microwave background (CMB), one can use the measured apparent size of the BAO as a 'standard ruler' in the late-time Universe. Used as a distance measure in this way, the BAO are one of the most powerful cosmological observables that can be derived from large-scale structure (LSS) surveys – making it possible to accurately reconstruct the geometry and expansion history of the Universe, while being remarkably robust to systematic errors. In combination with the CMB and other observations, precision BAO measurements are able to decisively answer fundamental questions about the nature of dark energy (especially its possible evolution with redshift), possible modifications to General Relativity, and the spatial curvature of the Universe.

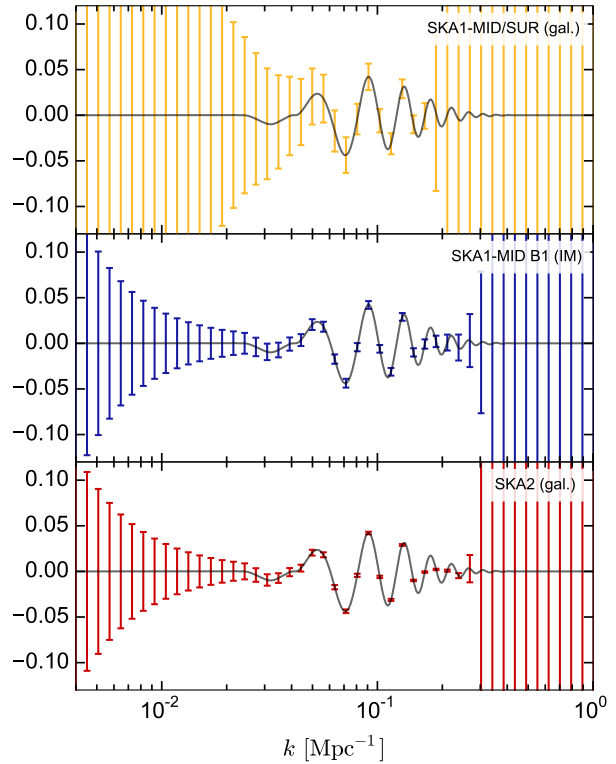

The SKA will be able to measure the BAO in two ways (Fig. 1). The first, using the intensity mapping (IM) technique, seeks to reconstruct the cosmological density field in three dimensions over unprecedentedly large volumes, using the 21cm spin-flip transition of neutral hydrogen to trace the density field in redshift space. The second is a more conventional galaxy redshift survey, which will detect many millions of discrete sources over a wide range of redshifts, with accurate redshift determinations coming from the 21cm line. Of the two, IM is the newer and less-tested method – galaxy surveys have been used to great effect as probes of LSS over the past several decades, whereas the first IM experiments are only just coming online, and are subject to a number of potential foregrounds and systematics that are yet to be fully quantified. Nevertheless, an IM survey is the best prospect for doing transformative cosmology with Phase 1 of the SKA – surpassing all existing BAO constraints at low redshift, and strongly complementing future surveys at higher redshift.

**Figure 1:** Constraints on the BAO 'wiggles', $f_{\rm BAO}(k) = P(k)/P_{\rm ref}(k) - 1$ (where $P_{\rm ref}(k)$ is a BAO-free reference spectrum), combined over all available redshift bins for three SKA galaxy/IM surveys. Full forecasts are given in Section 4.

In this chapter, we will review the physics of the BAO, its suitability for use as a standard ruler, and the current state of the art in BAO measurements from optical and Lyman-$\alpha$ surveys. We will then describe two possible methods for reconstructing the BAO with large surveys on the SKA, and present forecasts of the capabilities of SKA Phase 1 and 2 for each.

## 2. BAO as a standard ruler

In the early Universe, baryons and radiation were tightly coupled together via interactions with free electrons, by Coulomb attraction and Thomson scattering respectively. The tendency of overdensities in the baryon distribution to collapse under gravity was resisted by the increase in radiation pressure of photons collapsing along with them, setting up acoustic oscillations in the coupled photon-baryon plasma. The speed of sound in the fluid gives rise to a preferred scale – the *sound horizon*, $r_s$ – corresponding to the distance a sound wave could have travelled since the Big Bang. As the Universe cooled and the relative proportion of radiation to matter fell, Thomson scattering became inefficient and the photons and electrons decoupled, allowing the radiation to stream away. With the radiation pressure gone, the sound waves stalled, and the pattern of over- and under-densities that they had set-up were left frozen into the baryon distribution with a typical correlation length corresponding to the sound horizon at this time. Cold dark matter, being coupled to the baryonic matter gravitationally, was drawn into the same pattern of fluctuations, leaving a preferred distance scale in the total matter distribution.

One can measure this distance scale in a statistical manner by reconstructing the matter correlation function, $\xi(\mathbf{r}) = \langle \delta_M(\mathbf{x}) \delta_M(\mathbf{x}+\mathbf{r}) \rangle$, from surveys of large-scale structure using galaxies or other tracers of the matter density field. The BAO scale appears as a relatively broad 'bump' feature on $\sim 150$ Mpc scales, and so large survey volumes are needed to measure it well. One only measures a *relative* distance scale from the correlation function; retrieving physical distances depends on knowing the comoving sound horizon (i.e. the physical scale of the oscillations) as well. This can be calibrated from the CMB angular power spectrum, for example.

The BAO can be measured separately in the radial and tangential directions. If one makes no attempt to separate the two (i.e. by taking an angular average), the distance measured is $d(z) = r_s(z_d)/D_V(z)$, where $z_d$ is the redshift of the baryon drag epoch, and the *dilation scale* is given by

$$D_V(z) = \left( (1+z)^2 D_A^2(z) \frac{cz}{H(z)} \right)^{\frac{1}{3}}. \tag{2.1}$$

The expansion rate, $H(z)$, and angular diameter distance, $D_A(z)$, describe the geometry of the Universe, so measurements of $D_V(z)$ over a range of redshift can be used to constrain the evolution of the equation of state of dark energy, $w(z)$, for example. As shown by Eq. (2.1), however, one only measures a combination of $H(z)$ and $D_A(z)$, and so model degeneracies can arise. The degeneracy can be broken by separating the BAO in each direction; the quadrupole of the correlation function is sensitive to the combination

$$F(z) = (1+z)D_A(z)H(z)/c, \tag{2.2}$$

and the combination of $D_V$ and $F$ uniquely determines $D_A$ and $H$. In practise, one can separate out the tangential and radial directions by marginalising over the shape of the clustering pattern, without needing to understand the effects of redshift space distortions (Anderson et al. 2013).

### 2.1 Constraining dark energy with distance measurements

A measurement of $H(z)$ and $D_A(z)$ can be directly related to the dynamics and geometry of space time. Specifically, the Friedmann equations relate the expansion rate to the energy content and curvature of the universe via

$$H^2(z) = \frac{8\pi G}{3} \left[\rho_M(1+z)^3 + \rho_R(1+z)^4 + \rho_{\rm DE}(z)\right] - \kappa(1+z)^2$$

where $\rho_M$ and $\rho_R$ are the energy density in matter and radiation today, $\kappa$ is the curvature of space and $\rho_{\rm DE}(z)$ encapsulates all the unaccounted "dark" energy which may be contributing to the expansion of the universe. This dark energy can take a wide range of different forms: scalar fields (or quintessence) which are analogous to the Higgs field of the standard model (Copeland et al. 2006), gravitational degrees of freedom that are a remnant from higher dimensions, or even more radical modifications to Einstein's General Theory of Relativity (Clifton et al. 2012). Given its exotic nature and the fact that we cannot account for the dark energy with any of the other, known forms of energy, this is a unique opportunity to explore physics beyond the standard model and learn something qualitatively new about the fundamental laws of nature.

To learn about dark energy using BAO, we attempt to pin down its time evolution. Given that we are solely focusing on the dynamics and geometry of the background spacetime, $\rho_{\rm DE}(z)$ is completely determined by its value today, $\rho_{\rm DE}(0)$, and its equation of state, $w(z)$, such that $d\ln\rho_{\rm DE}/d\ln z = 3[1+w(z)]$. It is by characterising $w(z)$ that we hope to learn about the underlying physics that is driving the expansion of the Universe at late times. In theory $w(z)$ can have a wide range of behaviours, but in practice it is useful to consider simpler parametrizations. A particularly popular and useful parametrisation is to – effectively – Taylor expand the equation of state in terms of the scale factor $a \equiv (1+z)^{-1}$ so that

$$w \simeq w_0 + w_a(1-a), \quad (2.3)$$

where $w_0$ and $w_a$ are constants. While clearly oversimplifying the time evolution of $w(z)$ and in principle only accurate for $z \ll 1$, this parametrisation turns out to capture the behaviour of a very broad class of models for dark energy. Furthermore, it can be shown that, under general conditions, there are reasonably tight consistency relations between $w_0$ and $w_a$ that can be used to select out particular types of physical model. In this paper we will use $w_0$ and $w_a$ to quantify the precision of our constraints, although more general approaches have been used which involve decomposing the full history of $w(z)$ into suitable functional components (Zhao et al. 2014).

A precise measurement of $w_0$ and $w_a$ can lead to profound insights. Current constraints are weak – especially from $z \gtrsim 1$ – but are consistent with $w_0 = -1$ and $w_a = 0$, which is what one might expect from a cosmological constant, in itself quite a remarkable result. If future constraints show that $w_0 \neq -1$ and $w_a \neq 0$ then it would correspond to the discovery of a new degree of freedom in the Universe: a new fundamental field or modifications to general relativity on cosmological scales. If one were to find $w_0 < -1$ (that is, if the equation of state were to cross the "phantom divide" at late times (Caldwell et al. 2003)), then the consequences would be quite dramatic and one would have to consider a substantial revision of the current rules of field theory and gravitation.

## 2.2 Systematic effects and density field reconstruction

One of the major advantages of using the BAO to measure distances is their robustness to systematic effects. Since the distance information is primarily contained in the *location* of a feature in the correlation function (or equivalently the power spectrum), rather than its amplitude or detailed shape, there is little dependence on difficult-to-calibrate quantities like the overall normalisation of the matter power spectrum. Nevertheless, the non-linear growth of structure and scale-dependent effects can blur and bias the measured location of the BAO feature, so must be taken into account.

Consider how galaxies respond to the growth of large-scale structure, for example. As they fall toward their local clusters and superclusters, the large-scale galaxy pairs which encode the pristine baryon acoustic scale are shifted to smaller or larger separations, broadening the baryon acoustic peak and decreasing the accuracy with which it can be recovered. The technique of "density-field reconstruction" (Eisenstein et al. 2006) performs an approximate computation of these motions using the observed distribution of galaxies, and hence restores objects to the near-original position in the linear density field. This has the effect of sharpening the acoustic peak and significantly improving distance measurements.

The potential level of improvement depends on the number density of the tracer, as well as the individual noise of the realisation, but for sample variance-limited surveys can be a factor of $\sim 2$ (Seo & Eisenstein 2007). The reconstruction technique has been applied with success to BAO measurements in the largest galaxy redshift surveys at intermediate redshifts: the Sloan Digital Sky Survey (SDSS) Luminous Red Galaxy sample (Padmanabhan et al. 2012), the Baryon Oscillation Spectroscopic Survey (BOSS) (Anderson et al. 2013) and the WiggleZ Survey (Kazin et al. 2014).

Non-linear structure formation can also introduce scale-dependent effects. One typically thinks of the present-day density field becoming significantly non-linear only on scales below $\sim 30$ Mpc, far below the BAO scale at $\sim 150$ Mpc. Mode coupling on small scales, and the fact that the observed correlation function is a convolution over all Fourier modes, means that non-linear effects have an impact on considerably larger scales, however. Galaxy bias also affects the shifts introduced by non-linearities. While non-linear effects that change the shape of the clustering pattern can be marginalised without biasing the location of the BAO feature, the combination of all effects can introduce a modest (but non-negligible) $\sim 0.1 - 0.2\%$ bias after reconstruction (Seo et al. 2010; Mehta et al. 2011).

## 2.3 Previous measurements

Since the first significant detection of the baryon acoustic peak in the SDSS Luminous Red Galaxy (LRG) sample by (Eisenstein et al. 2005), BAO measurements have improved along with the increasing volume mapped by galaxy redshift surveys, and now provide per-cent level distance and expansion measurements at various cosmic epochs. The most precise current measurements derive from BOSS (DR11, (Anderson et al. 2013)), which now offers 2% and 1% distance measurements at $z = 0.32$ and $z = 0.57$ respectively, using nearly one million galaxies covering 8,500 deg$^2$ of sky and a volume of 13 Gpc$^3$. BOSS has also measured BAOs in the structure of the Lyman-$\alpha$ forest on the sightlines to quasars, providing a 2% distance measurement at $z = 2.34$ (Delubac et al. 2014).

Measurements at lower precision have previously been reported by the WiggleZ Dark Energy Survey ($\sim 4\%$ measurements in two independent bins at $z = 0.44$ and $z = 0.73$, (Kazin et al. 2014)), the 6-degree Field Galaxy Survey (4.5% measurement in the local Universe at $z = 0.1$, effectively serving as an independent determination of $H_0$ (Beutler et al. 2011)), and the final SDSS LRG sample (Padmanabhan et al. 2012). BAOs have also been detected in photometric redshift surveys at intermediate redshifts (Seo et al. 2012).

Upcoming optical and near-infrared redshift surveys enabling BAO measurements include the extended BOSS project (eBOSS), which will provide precision distance measurements in the range $z < 1$, the Hobby-Eberly Telescope Dark Energy Experiment (HETDEX), targeting $2 < z < 4$, the Euclid and WFIRST satellites, and proposed ground-based projects such as the Dark Energy Survey Instrument (DESI) and the VISTA/4MOST telescope. Ref. (Font-Ribera et al. 2013) describe how a combination of these facilities should provide percent-level distance constraints over the range $0 < z < 4$. In this chapter we discuss whether SKA surveys are able to compete in this landscape.

## 3. Experiment and survey design

There are two main methods that can be used to measure the BAO with the SKA: a galaxy redshift survey and an intensity mapping survey. In this section, we describe the relative merits of the two methods, and define baseline specifications for both types of survey on the Phase 1 and 2 configurations.

### 3.1 HI galaxy redshift survey

Galaxies are biased tracers of the cosmological density field. By detecting many individual galaxies and measuring their redshifts, one can constrain the matter correlation function (or equivalently, the power spectrum). As discussed above, redshift surveys at optical wavelengths have been used to great effect for cosmology, and with the SKA one will be able to do the same in the radio.

Most important is the choice of target galaxy population. In particular, one requires galaxies with an easily measurable emission/absorption feature for measuring redshifts – the most appropriate for the SKA being HI (Santos et al. 2014b). The high spectral resolution of SKA receiver systems rivals the redshift resolution of optical spectroscopic surveys, and their large bandwidths allow a wide redshift range to be covered in principle. The Phase 1 arrays have insufficient sensitivity to yield competitively-sized samples of HI galaxies, however – even with optimistic assumptions, a 10,000 hour survey over 5,000 deg$^2$ with SKA1-MID or SUR will achieve an RMS flux sensitivity of $S_{\rm rms} \approx 70 - 100$ $\mu$Jy, equating to roughly $5 \times 10^6$ galaxies out to $z \approx 0.5$ for a $5\sigma$ detection threshold. This is worse than the expected yield from the full 10,000 deg$^2$ BOSS survey, which will be completed long before Phase 1 sees first light. SKA2, on the other hand, will be far more sensitive, reaching $S_{\rm rms} \approx 5$ $\mu$Jy for a 10,000 hour survey over 30,000 deg$^2$, even with a more stringent $10\sigma$ threshold. The expected yield for such a survey is $\sim 10^9$ galaxies between $0.18 < z < 1.84$, far surpassing any planned optical or near-infrared survey for $z \lesssim 1.4$. The predicted number density and bias of HI galaxies for SKA1-MID (including MeerKAT dishes), SKA1-SUR (including ASKAP dishes), and SKA2 are given in Table 1; more details on the sensitivity calculation are given in Santos et al. (2014b).

There are a number of potential systematic effects that can affect galaxy redshift surveys. In large optical galaxy surveys at least, stars are a major contaminant – while one can distinguish stars from galaxies by their colour, bright stars effectively mask galaxies behind them, leading to a complicated angular selection function on the sky. This is less of an issue in the radio, although other contaminants, such as diffuse galactic synchrotron emission and non-galaxy point sources, can also cause problems for the source-finding algorithms used to compile the galaxy catalogue. Source-finding in radio data is also made more challenging by needing to search through 3D datacubes resolved in angle and frequency, rather than just sets of discrete 2D images.

| SKA1-MID | | |
|---|---|---|
| $z_c$ | $n(z)$ [Mpc$^{-3}$] | $b(z)$ |
| 0.05 | $2.92 \times 10^{-2}$ | 0.678 |
| 0.15 | $6.74 \times 10^{-3}$ | 0.727 |
| 0.25 | $1.71 \times 10^{-3}$ | 0.802 |
| 0.35 | $4.64 \times 10^{-4}$ | 0.886 |
| 0.45 | $1.36 \times 10^{-4}$ | 0.975 |

| SKA1-SUR | | |
|---|---|---|
| $z_c$ | $n(z)$ [Mpc$^{-3}$] | $b(z)$ |
| 0.05 | $4.14 \times 10^{-2}$ | 0.664 |
| 0.15 | $8.00 \times 10^{-3}$ | 0.724 |
| 0.25 | $1.56 \times 10^{-3}$ | 0.802 |
| 0.35 | $3.39 \times 10^{-4}$ | 0.890 |
| 0.45 | $7.86 \times 10^{-5}$ | 0.989 |
| 0.55 | $1.91 \times 10^{-5}$ | 1.099 |
| 0.65 | $4.77 \times 10^{-6}$ | 1.221 |
| 0.75 | $1.23 \times 10^{-6}$ | 1.357 |
| 0.85 | $3.21 \times 10^{-7}$ | 1.507 |

| SKA2 | | |
|---|---|---|
| $z_c$ | $n(z)$ [Mpc$^{-3}$] | $b(z)$ |
| 0.23 | $4.43 \times 10^{-2}$ | 0.713 |
| 0.33 | $2.73 \times 10^{-2}$ | 0.772 |
| 0.43 | $1.65 \times 10^{-2}$ | 0.837 |
| 0.53 | $9.89 \times 10^{-3}$ | 0.907 |
| 0.63 | $5.88 \times 10^{-3}$ | 0.983 |
| 0.73 | $3.48 \times 10^{-3}$ | 1.066 |
| 0.83 | $2.05 \times 10^{-3}$ | 1.156 |
| 0.93 | $1.21 \times 10^{-3}$ | 1.254 |
| 1.03 | $7.06 \times 10^{-4}$ | 1.360 |
| 1.13 | $4.11 \times 10^{-4}$ | 1.475 |
| 1.23 | $2.39 \times 10^{-4}$ | 1.600 |
| 1.33 | $1.39 \times 10^{-4}$ | 1.735 |
| 1.43 | $7.99 \times 10^{-5}$ | 1.882 |
| 1.53 | $4.60 \times 10^{-5}$ | 2.041 |
| 1.63 | $2.64 \times 10^{-5}$ | 2.214 |
| 1.73 | $1.51 \times 10^{-5}$ | 2.402 |
| 1.81 | $9.66 \times 10^{-6}$ | 2.566 |

**Table 1:** Predicted number density and bias of HI galaxies as a function of redshift, for SKA1-MID (including MeerKAT, Band 2), SUR (including ASKAP), and SKA2 ($\Delta z = 0.1$ bins). All assume 10,000 hour surveys, over 5,000 deg$^2$ (SKA1-MID and SUR) and 30,000 deg$^2$ (SKA2). The detection thresholds are chosen to be $5\sigma$, $5\sigma$, and $10\sigma$ respectively. See Santos et al. (2014b) for more details.

One should also be careful of source evolution effects. For example, the luminosity function of the tracer population is expected to change with redshift, modifying the number of galaxies that can be detected. Depending on the tracer, this limits the useful redshift range of a survey, and complicates its selection function. This evolution is commonly parametrised as an evolution of the galaxy bias, $b(z)$, and the principal effect on the BAO is to change the shot noise by changing the effective galaxy number density, $n(z)$. It is also possible that the bias can become scale-dependent, however, which introduces the possibility of systematically biasing the BAO scale.

As discussed in Section 2.2, redshift space distortions and effects on non-linear scales also affect the measured power spectrum. A simple model for the impact of RSDs and non-linearities on the galaxy power spectrum is given by (Kasier 1987; Seo & Eisenstein 2007),

$$P_{\rm tot}(\mathbf{k}) = (b(z) + f(z)\mu^2)^2 e^{-\frac{1}{2}k^2 \sigma_{\rm NL}^2(z,\mu)} P(k,z), \tag{3.1}$$

$$\sigma_{\rm NL}(z,\mu) = \sigma_{\rm NL} D(z) \left(1 + f(z)\mu^2[2 + f(z)]\right)^{\frac{1}{2}} \tag{3.2}$$

where $D(z)$ is the growth factor, $f(z) = d\log D/d\log a$ is the linear growth rate of structure, and $P(k,z)$ is the isotropic matter power spectrum. The leading term of Eq. (3.1) is the anisotropy due to RSDs ($\mu \equiv \cos\theta$), and the exponential term models the "washing-out" of redshift information due to incoherent non-linear velocities, characterised by the velocity dispersion, $\sigma_{\rm NL}$. These effects are discussed in more detail in Raccanelli et al. (2014).

### 3.2 Intensity mapping survey

If one is only interested in measuring the matter distribution on large scales, there is not strictly any need to resolve individual galaxies. Instead, one can perform a survey with relatively low angular resolution that detects only the integrated intensity from many unresolved sources. This is called intensity mapping (IM), and allows extremely large volumes to be surveyed very efficiently.

The SKA will be capable of performing large IM surveys over $0 \lesssim z \lesssim 3$ using the redshifted neutral hydrogen 21cm emission line (Santos et al. 2014a). Neutral hydrogen is ubiquitous in the late universe, residing principally in dense regions inside galaxies that are shielded from the ionising UV background, and the 21cm line is narrow and relatively unaffected by absorption or contamination by other lines. It is thus an excellent tracer of the matter density field in redshift space. The background brightness temperature of the HI emission is (Bull et al. 2014)

$$\overline{T}_b = \frac{3}{32\pi} \frac{hc^3 A_{10}}{k_B m_p v_{\rm HI}^2} \frac{(1+z)^2}{H(z)} \Omega_{\rm HI}(z) \rho_{c,0}, \qquad (3.3)$$

where $A_{10}$ is the Einstein coefficient for emission, and $\Omega_{\rm HI}(z)$ is the fraction of the critical density today, $\rho_{c,0}$, in neutral hydrogen. Assuming that HI is a linearly-biased tracer of the total matter density field, we can relate fluctuations in the brightness temperature to the (Fourier-transformed) redshift-space matter density perturbation,

$$\delta T_b = \overline{T}_b \delta_{\rm HI}(\mathbf{k}) = \overline{T}_b (b_{\rm HI} + f\mu^2) e^{-\frac{1}{4}k^2 \sigma_{\rm NL}^2(z,\mu)} \delta_M(k), \qquad (3.4)$$

where the anisotropic terms are due to RSDs and non-linear growth (Eq. 3.1). One can then measure the 3D redshift-space matter power spectrum, $\langle \delta_M(\mathbf{k}) \delta_M^*(\mathbf{k}') \rangle = (2\pi)^3 \delta^{(3)}(\mathbf{k} - \mathbf{k}') P(\mathbf{k})$, by mapping out the brightness temperature distribution in angle and frequency, $\nu = \nu_{\rm HI}/(1+z)$.

|  | SKA1-MID | | SKA1-SUR | |
|---|---|---|---|---|
|  | Band 1 | Band 2 | Band 1 | Band 2 |
| $T_{\rm inst}$ [K] | 28 | 20 | 50 | 30 |
| $z_{\rm min}$ | 0.35 | 0.00 | 0.58 | 0.00 |
| $z_{\rm max}$ | 3.05 | 0.49 | 3.05 | 1.18 |
| $\nu_{\rm min}$ [MHz] | 350 | 950 | 350 | 650 |
| $\nu_{\rm max}$ [MHz] | 1050 | 1760 | 900 | 1670 |
| $D_{\rm dish}$ [m] | 15 | 15 | 12 | 12 |
| $\delta\nu$ [kHz] | 50 | 50 | 50 | 50 |
| $\Omega_{\rm sur}$ [$10^3$ deg$^2$] | 25 | 25 | 25 | 25 |
| $N_{\rm dish} \times N_{\rm beam}$ | $254 \times 1$ | $254 \times 1$ | $60 \times 36$ | $60 \times 36$ |

**Table 2:** Baseline IM survey specifications for the Phase 1 SKA configurations. SUR is equipped with PAFs; the assumed scaling of the PAF FOV with frequency is explained in Santos et al. (2014b).

Example survey specifications for Phase 1 of the SKA are shown in Table 2. One of the key decisions in designing an IM survey for the SKA will be whether to use the array in autocorrelation (single-dish) or cross-correlation (interferometer) mode. As one is interested in mapping out extended structure on comparatively large angular scales rather than detecting individual galaxies that subtend only small angles, the angular sensitivity is an important factor. Roughly speaking, single-dish experiments are sensitive to angular scales between the field of view (FOV) of a single dish and the full area of the survey, whereas interferometers can only see between the scales corresponding to their minimum and maximum baseline lengths. Since the minimum baseline can be no smaller than the diameter of a single dish, the maximum possible angular scale that interferometers are sensitive to is set by the single-dish FOV (i.e. the primary beam). The two modes are therefore in some sense complementary (Fig. 2).

For the $\sim 15$m dishes of the SKA, the physical scales corresponding to the BAO are best matched to an autocorrelation survey for $z \lesssim 1$, and an interferometric survey at higher $z$. Having the BAO features fall somewhere within an instrument's resolution window is only a minimum requirement, however – ideally, one should maximise its sensitivity at the relevant angular scales too. This is trivial in the single-dish case, which has an essentially flat response over its full range of angular scales, but for interferometers there is a strong dependence on the detailed baseline distribution. For IM with the SKA, a high density of short baselines is optimal, suggesting an array configuration with highly-clustered stations. For the currently-proposed SKA1 configurations, however, the density of short baselines is not high enough, and autocorrelation mode always wins out on sensitivity at low redshift. An interferometric survey would be better suited to BAO detections at high redshift, $z > 1.5$, but this is a less interesting range for constraining dark energy. A proposed dense aperture array component of SKA2 would likely be better suited to intensity mapping; aperture arrays have the large FOV and high density of short baselines necessary for sensitivity to the angular scales of the BAO at lower $z$.

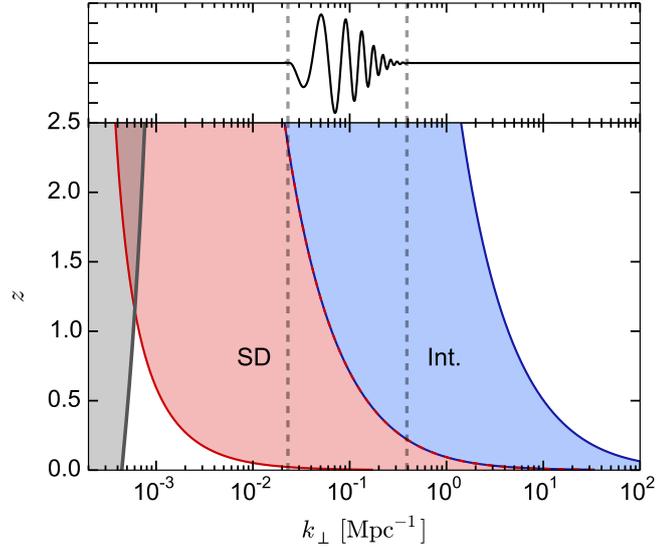

**Figure 2:** Maximum and minimum transverse (angular) scales that can be probed as a function of redshift in single-dish (autocorrelation) and interferometer modes, for a representative SKA array (15m dishes with baselines restricted to a 1 km core, and 25,000 deg$^2$ survey area). The BAO (dashed vertical lines) and comoving horizon scale (grey band) are shown for comparison. (N.B. the array will also be sensitive to structure in the radial (frequency) direction.)

Though promising, the intensity mapping methodology is not yet mature, and it remains to be seen whether a number of potentially serious issues can be overcome. The most obvious of these is the presence of foreground emission from the galaxy and extragalactic point sources, which strongly contaminates the HI signal. Fortunately, despite being several orders of magnitude brighter, most of the foregrounds are spectrally smooth and so can be readily distinguished from all but the largest-scale modes of the fluctuating cosmological signal. More insidious is the leakage of polarised foreground emission into the total intensity channel; this is not spectrally smooth because of frequency-dependent Faraday rotation caused by the partially-incoherent magnetic field of the galaxy (Alonso et al. 2014). The issue of foreground contamination is explored in more detail in Wolz et al. (2014).

Another significant issue for single-dish surveys in particular is correlated ("$1/f$") noise. A well-known problem in CMB analysis, correlated noise from the receiver system and elsewhere dominates the uncorrelated white noise component on long timescales. This causes the signal-to-noise ratio to grow more slowly as a function of integration time, and introduces striping artefacts into maps of the emission, seriously hindering the recovery of large-scale modes of the signal. For the CMB, this is handled by using receivers with low knee frequencies ($< 1$ Hz), scanning rapidly across the sky, and filtering out timescales longer than $1/f_{\rm knee}$ from the time-ordered data. An SKA autocorrelation survey would have to adopt similar strategies to make a wide-angle survey viable. Finally, ground pickup/spillover is also an issue for autocorrelation surveys, and must be suppressed (e.g. with ground shielding), or mapped out and subtracted.

## 4. Forecasts for the SKA

In this section we present forecasts for BAO distance measurements – and the resulting constraints on cosmological parameters – for several SKA configurations, and compare them with what will be possible with other methods in around the same timeframe. The forecasts are based on the Fisher forecasting formalism developed in (Bull et al. 2014), using the number counts from Table 1 (galaxy survey), and experimental specifications set out in Table 2 (IM survey). Complementary forecasts for redshift space distortions are given in Raccanelli et al. (2014).

While there are a number of possible combinations of arrays, frequency bands, and survey modes for both Phase 1 and 2, for compactness we have chosen to concentrate on the best-performing configurations. For example, for Phase 1 we neglect Band 1 of SUR and MID for galaxy surveys, as they will detect comparatively few HI galaxies, and we consider only autocorrelation mode for intensity mapping, since the angular resolution of an interferometric IM survey on the SKA is poorly matched to the BAO for $z \lesssim 1$. We present forecasts only for a galaxy survey for SKA2, although it should be noted that an IM survey on a mid-frequency aperture array should be able to provide similar constraints on the BAO out to $z \approx 2$.

### 4.1 Fisher forecasting

To accurately characterise the expected performance of a given experiment, one would ideally perform a full simulation, incorporating a variety of potential systematic and instrumental effects, and running the simulated data through the actual analysis pipeline. This is computationally intensive, and detailed aspects of the hardware and signal may not yet be known, as is the case here. Fisher forecasting takes a simpler approach, instead using the expected properties of the signal and noise for an experiment to derive a Gaussian approximation to the likelihood for a set of parameters to be measured. Though clearly not definitive, Fisher forecasts at least take into account important effects like correlations between parameters, and are sufficiently accurate for understanding the relative performance of different experiments.

The key procedure in Fisher forecasting is to construct the Fisher matrix, $F$, for a set of parameters, $\theta$. When inverted, this yields an estimate of the covariance matrix for the Gaussianised likelihood. The Fisher matrices for both IM and galaxy redshift surveys can be written in the form

$$F_{ij} = \frac{1}{2} \int \frac{d^3k}{(2\pi)^3} V_{\text{eff}}(\mathbf{k}) \left[ \frac{\partial}{\partial \theta_j} \log C^T \cdot \frac{\partial}{\partial \theta_j} \log C^T \right], \qquad (4.1)$$

where $C^T = C^S + C^N$ is the total covariance of the measured signal, consisting of the true signal (S) and noise (N), and $V_{\text{eff}}(\mathbf{k}) = V_{\text{phys}} \left( C^S/C^T \right)^2$ is the effective volume of the survey (which covers a physical volume $V_{\text{phys}}$). Only $C^S$ is a function of the cosmological parameters of interest, $\{\theta\}$. For a galaxy survey, the total signal is the galaxy power spectrum plus shot noise, $C^T = P_{\text{tot}}(\mathbf{k}) + 1/n(z)$. For intensity mapping, the expression is more complicated: $C^S \propto T_b^2 P_{\text{tot}}(\mathbf{k})$ and $C^N \propto T_{\text{sys}}^2 S_{\text{area}}/t_{\text{tot}} B(\mathbf{k})$, where $B$ is a window function that depends on the angular and frequency resolution of the radio telescope (full expressions for both interferometer and autocorrelation experiments are given in Bull et al. (2014)). We have neglected the effects of foreground subtraction here.

To calculate the Fisher matrix, we must specify a fiducial cosmology and a set of parameters to be measured (including nuisance parameters). We adopt the Planck best-fit ΛCDM model (Planck Collaboration 2013),

$$h = 0.67, \ \Omega_\Lambda = 0.684, \ \Omega_K = 0, \ \Omega_b = 0.049, w = -1, \ n_s = 0.962, \ \sigma_8 = 0.834, \ N_{\text{eff}} = 3.046, \qquad (4.2)$$

and in the first instance forecast for the parameters

$$\left\{ \alpha_\perp = D_A(z)|_{\text{fid.}}/D_A(z), \alpha_\parallel = H(z)/H(z)|_{\text{fid.}}, \sigma_8 f(z), \sigma_8 b(z), n_s, \sigma_{\text{NL}} \right\}, \qquad (4.3)$$

where 'fid.' denotes evaluation in the fixed fiducial cosmology. Since we are only interested in the BAO here, we discard information on the distance parameters $\{\alpha_\perp, \alpha_\parallel\}$ from all other sources. In practise, this means that we only keep the terms with derivatives of the BAO part of the power spectrum with respect to $\alpha$, i.e. $\partial f_{\text{BAO}}(k)/\partial \alpha$, where we have split the isotropic power spectrum into a smooth part and an oscillating part, $P(k) = [1 + f_{\text{BAO}}(k)] P_{\text{smooth}}(k)$. We also assume that no reconstruction of the density field is performed.

Finally, one can project from this set of parameters into another that corresponds directly to the parameters of a cosmological model. We will consider the projection of the distance/growth functions, $\{\alpha_\perp, \alpha_\parallel, f\}$, into the parameters

$$\{h, n_s, \Omega_\Lambda, \Omega_K, w_0, w_a, \sigma_{\text{NL}}, \sigma_8, b(z)\}. \qquad (4.4)$$

This corresponds to fitting a cosmological model to the distance measurements extracted from the BAO, and the growth rate measured from the anisotropy of the correlation function.

### 4.2 BAO forecasts for the SKA

Predicted constraints on the BAO feature in the power spectrum for three different SKA surveys were shown in Fig. 1, as a function of scale. For Phase 1, it is clear that an intensity mapping survey will have much better overall sensitivity to the BAO than a galaxy survey when constraints are combined over the full redshift range. A galaxy survey with SKA2 will far surpass both of these (subject to various systematic effects).

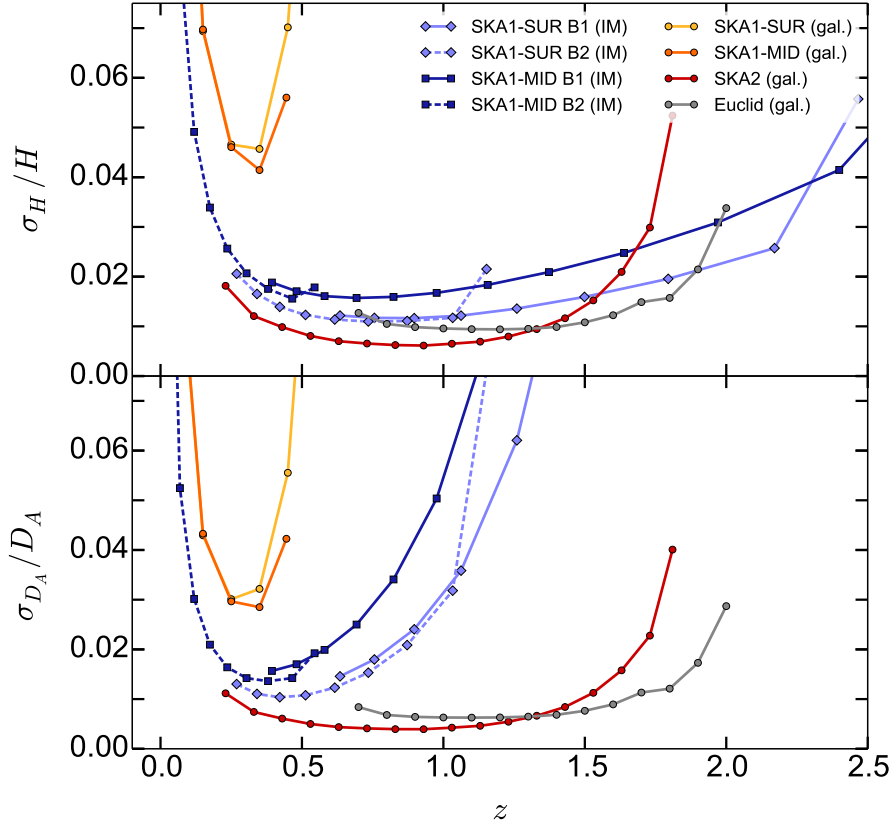

**Figure 3:** BAO-only fractional constraints (68% CL) on the expansion rate (top panel) and angular diameter distance (bottom panel) for SKA IM and galaxy redshift surveys. The bias has been marginalised as a free parameter in each redshift bin. The degradation of the $D_A$ constraints at higher $z$ for the IM experiments is caused by their falling angular resolution (c.f. Fig. 2).

Fig. 3 shows forecast constraints on the expansion rate and angular diameter distance for the proposed SKA surveys, in bins of width $\Delta z = 0.1$ (galaxy surveys) or $\Delta \nu = 60$ MHz (IM surveys). Forecasts for a Euclid galaxy survey are also shown for comparison, based on the predicted number counts and bias in Amendola et al. (2013). For Phase 1, a galaxy survey will not be competitive with other BAO measurements owing to insufficient sensitivity. An IM autocorrelation survey will be significantly more powerful, providing constraints on $H(z)$ that are similar to the (sample variance-limited) Euclid experiment, but over a significantly wider redshift range – Band 2 of both SUR and MID will yield sub-2/3% constraints beyond $z \simeq 2$. Constraints on the angular diameter distance will be considerably worse at higher redshift due to the limited angular resolution in autocorrelation mode, however (c.f. Fig. 2). A galaxy survey with SKA2 will be sample variance-limited over 30,000 deg² for $0.4 \lesssim z \lesssim 1.3$, surpassing all other planned surveys over that range.[1]

---

[1] An IM survey on a Phase 2 dense mid-frequency aperture array operating from 450 MHz upwards could provide similarly tight constraints on $D_A$ and $H$ all the way out to $z = 2$ if a large enough collecting area could be built.

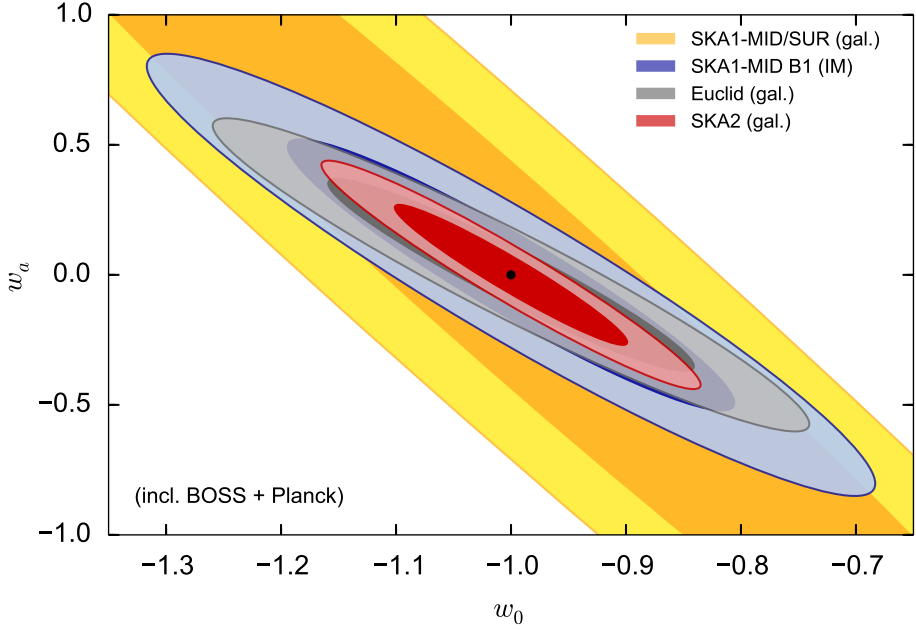

**Figure 4:** Forecast constraints on $(w_0, w_a)$ for several SKA configurations and Euclid, in combination with Planck and BOSS. All other parameters have been marginalised, including $\Omega_K$, and the bias is free per $z$ bin.

Expected constraints on the dark energy equation of state parameters (Eq. 2.3) are shown in Fig. 4 for a subset of these configurations, combined with forecast Planck CMB and BOSS low redshift BAO constraints.[2] Forecasts for a future Euclid galaxy survey are shown for comparison, also with Planck+BOSS included. While unable to match Euclid on raw figure of merit[3] (FOM = 69, versus 129 for Euclid), the Phase 1 IM survey has a highly complementary redshift range and wide survey area (with potentially $\sim 100\%$ overlap), and should be completed in the early 2020's, making it the low redshift dataset of choice for joint analyses with Euclid and other high redshift experiments like WFIRST and DESI. Combining surveys will be important for pinning-down the nature of dark energy in the medium term, as "Stage IV" galaxy surveys may not offer sufficient precision on their own to discriminate between most dark energy models (Marsh et al. 2014). In the longer term, the galaxy survey with SKA2 will be able to achieve a substantially larger FOM of around 310.

## 5. Conclusions

The Baryon Acoustic Oscillations imprint a distance scale into the distribution of matter on large scales that can be used to constrain the expansion history and geometry of the Universe. In turn, this can be used to constrain key cosmological parameters, potentially allowing us to answer fundamental questions about the nature of dark energy such as whether it evolves with time. The BAO are remarkably robust to systematic effects, and have been successfully measured to high precision by a number of optical galaxy redshift surveys and Lyman-$\alpha$ forest observations.

---

[2] We have assumed that the BOSS sample is statistically independent from the surveys that we combine it with.

[3] The dark energy figure of merit is defined as FOM $= 1/\sqrt{\det(F^{-1}|_{w_0,w_a})}$ (Albrecht et al. 2006).

The SKA will be able to measure the BAO through two different types of survey: HI galaxy redshift surveys, where millions of individual galaxies are detected and their redshifts measured from the HI emission line; and HI intensity mapping surveys, where the integrated HI emission from many unresolved galaxies is used to reconstruct the cosmological density field on large scales. Redshift surveys are a tried and tested technique, but require high sensitivity to capture enough galaxies. Intensity mapping has yet to mature, but is potentially a much more efficient way of detecting the BAO. Both suffer from a number of potential systematics, such as those associated with the non-linear evolution of the cosmological density field on small scales.

Phase 1 of the SKA will be able to produce competitive constraints on the (mostly radial) BAO at redshifts relevant for dark energy if a large IM autocorrelation survey can be performed. The Phase 1 arrays lack the sensitivity to detect enough galaxies to produce interesting constraints from a redshift survey, and there are an insufficient number of short baselines to make an interferometric IM survey worthwhile (unless higher redshifts are targeted). Note that an IM autocorrelation survey on an early deployment (pre-Phase 1) system may also be able to produce useful constraints on the expansion rate so long as sufficient survey time ($\sim$10,000 hours) can be obtained.

SKA2, on the other hand, will have the sensitivity to produce an immense galaxy redshift survey over almost ¾ of the sky, surpassing all other planned BAO measurements at $0.4 \lesssim z \lesssim 1.3$. This should allow it to pin down the equation of state of dark energy with unprecedented precision, assuming various systematic effects can be overcome. A mid-frequency aperture array could achieve similar precision at higher redshift (out to $z \simeq 2$) if a sufficiently large and dense collecting area can be used, but we have not considered this possibility in detail here.

Finally, while we have concentrated on the BAO as the most robust distance measure, redshift space distortions and even the overall shape of the power spectrum contain a great deal of extra information that can be used to constrain dark energy. In this sense, the forecasts here represent the most conservative estimates of the cosmological constraints that can be achieved with the SKA. If sufficient control over systematics can be achieved, considerably tighter measurements of $w(z)$ and the growth of structure can be expected by using these other probes too (Raccanelli et al. 2014).

*Acknowledgements* — PB is supported by European Research Council grant StG2010-257080. AR is supported by the Templeton Foundation. Part of the research described in this paper was carried out at the Jet Propulsion Laboratory, California Institute of Technology, under a contract with the National Aeronautics and Space Administration.